\documentclass[12pt]{amsart}

\usepackage{amsfonts,amsmath,amssymb,amsthm}
\usepackage{setspace,graphicx}
\usepackage{sgame}
\usepackage{natbib}
\usepackage{bm}
\usepackage{multirow}
\usepackage{ifthen}
\usepackage{verbatim}
\usepackage{caption}
\usepackage{enumitem}
\usepackage{threeparttable}
\usepackage{array}
\newcolumntype{L}[1]{>{\raggedright\let\newline\\\arraybackslash\hspace{0pt}}m{#1}}
\newcolumntype{C}[1]{>{\centering\let\newline\\\arraybackslash\hspace{0pt}}m{#1}}
\newcolumntype{R}[1]{>{\raggedleft\let\newline\\\arraybackslash\hspace{0pt}}m{#1}}

\usepackage{textcase}
\usepackage[T1]{fontenc}
\usepackage{fouriernc}
\makeatletter
\renewcommand\section{\@startsection{section}{1}{0mm}{-\baselineskip}{0.5\baselineskip}{\normalfont\centering\scshape}}
\renewcommand\subsection{\@startsection{subsection}{2}{0mm}{-\baselineskip}{0.5\baselineskip}{\normalfont\centering\itshape}}
\renewcommand\subsubsection{\@startsection{subsubsection}{3}{0mm}{-\baselineskip}{0.5\baselineskip}{\normalfont\centering\itshape}}
\renewcommand*{\@seccntformat}[1]{\csname the#1\endcsname\ \ }
\makeatother

\theoremstyle{plain}
\theoremstyle{definition}

  \newtheorem{corollary}{Corollary}
  \newtheorem{definition}{Definition}

  \newtheorem{lemma}{Lemma}
  \newtheorem{proposition}{Proposition}
  \newtheorem{remark}{Remark}
\theoremstyle{remark}

\renewcommand{\thesection}{\textbf{\Roman{section}.}}
\renewcommand{\thesubsection}{\Alph{subsection}.}

\renewcommand{\thefigure}{\Roman{figure}}
\renewcommand{\thetable}{\Roman{table}}

\makeatletter
\def\thickhline{%
  \noalign{\ifnum0=`}\fi\hrule \@height \thickarrayrulewidth \futurelet
   \reserved@a\@xthickhline}
\def\@xthickhline{\ifx\reserved@a\thickhline
               \vskip\doublerulesep
               \vskip-\thickarrayrulewidth
             \fi
      \ifnum0=`{\fi}}
\makeatother

\newlength{\thickarrayrulewidth}
\begin{document}

\thispagestyle{empty}
\title[Persuasion in the Long Run: When history matters]{Persuasion in the Long Run: When history matters}
\author[Hyeonggyun Ko]{Hyeonggyun Ko$^{*}$}
\thanks{$^{*}$The Ohio State University, Department of Economics. Email: ko.358@osu.edu.\\
I would like to thank Yaron Azrieli, P.J. Healy, James Peck, Huanxing Yang, and the participants of the OSU Theory/Experimental Reading Group, the Midwest International Trade and Theory Conference in Atlanta, and the 36th Stony Brook International Conference on Game Theory for their valuable comments and suggestions. All remaining errors are my own.
}
\singlespacing
\begin{abstract}

We study a long-run persuasion problem where a long-lived Sender repeatedly interacts with a sequence of short-lived Receivers who may adopt a misspecified model for belief updating. The Sender commits to a stationary information structure, but suspicious Receivers compare it to an uninformative alternative and may switch based on the Bayes factor rule. We characterize when the one-shot Bayesian Persuasion-optimal (BP-optimal) structure remains optimal in the long run despite this switching risk. 
In particular, when Receivers cannot infer the state from the Sender's preferred action, they never switch, and the BP-optimal structure maximizes the Sender's lifetime utility. In contrast, when such inference is possible, full disclosure may outperform BP-optimal. Our findings highlight the strategic challenges of information design when the Receiver’s interpretation of signals evolves over time.

\bigskip


\bigskip

  
\end{abstract}
\maketitle

\vspace{0.2in}

\begin{center}\Large{\textbf{\texttt{\today}}}\end{center}
\vspace{0.2in}
\bigskip

\onehalfspacing
\newpage

\section{\textbf{Introduction}}\label{sec:introduction}\

\indent When a long-lived Sender (``he'') repeatedly attempts to persuade a sequence of short-lived Receivers (each ``she'') to act in his preferred way, identical signals may not induce the same response in every period. For instance, consider a seller (Sender) trying to persuade a sequence of buyers (Receivers) to purchase a product each period through a test that signals its quality. Initially, buyers trust the test and buy the product upon receiving positive signals. However, if repeated purchases result in poor quality, their trust in the test may erode. As a result, even when receiving the same positive signal in later periods, they may choose not to purchase.

Motivated by this example, we study the optimal information disclosure policy of the long-lived Sender. In our model, the Sender commits to a stationary information structure, but Receivers do not fully trust it and compare the Sender's chosen structure with an alternative. 
Initially, Receivers adopt the Sender's structure. However, they may switch to the alternative if the alternative better explains the observed data, based on a Bayes factor rule \citep{Ba2023}.

We focus on an alternative structure that is (i) completely uninformative and (ii) generates the same signal distribution as the Sender's information structure.
The first feature captures the possibility that Receivers may lose trust in the informativeness of the signals and ignore them altogether.
The second feature reflects an environment where Receivers observe signal distributions but cannot verify the true underlying information structure (\citealp{lin2024credible}), leading to an alternative that mimics the same signal distribution as the Sender's structure.\footnote{The Receivers' switching between information structures depends critically on the nature of the alternative model, which may depend on the source of the Receiver's doubt about the Sender. Instead of the alternative considered in our model, one may consider other alternatives. For instance, Receivers may question how accurately the Sender transmits signals according to the intended information policy, leading them to consider a garbled alternative (\citealp{le2019persuasion}; \citealp{tsakas2021noisy}). Alternatively, if Receivers suspect the Sender's commitment, an alternative model may be akin to the Sender's profitable deviation in the cheap talk framework (\citealp{crawford1982strategic}).}

Unlike in the one-shot Bayesian Persuasion benchmark (\citealp{kamenica2011bayesian}; henceforth, BP), the Sender in our model must consider not only the period expected utility under his chosen information structure but also the risk of Receivers switching to the alternative and not responding to signals. We analyze the Sender's long-run persuasion problem in a binary-state, binary-action environment and ask: When is the optimal information structure in BP (BP-optimal) optimal in the long run? When do other structures outperform it?


The answers hinge on what Receivers observe each period: the history of signals, actions, and their period utilities up to that period. If the chosen action yields the same utility across states (non-revealing), Receivers do not learn about the realized state. Conversely, if their payoffs differ (revealing), Receivers can infer the realized state, gradually acquiring more knowledge about the underlying information structure.\footnote{For instance, in the seller-buyer example, purchasing the product is revealing since buyers' utility depends on the product's quality. In contrast, not purchasing is non-revealing, as it yields the same utility regardless of the quality.}

When every action is revealing, the Sender's chosen information structure will almost surely be adopted in the long run. Consequently, as the Sender becomes more patient, the BP-optimal information structure can outperform any other information structure in the long run. (\textbf{Proposition~\ref{prop:revealing}})


When one action is non-revealing, and the Sender has a preferred action, whether the BP-optimal structure retains its long-run optimality depends critically on whether the Sender's preferred action is revealing or non-revealing. \textbf{Proposition~\ref{prop:speed}} establishes that if the Sender's preferred action is non-revealing, Receivers never switch from the BP-optimal information structure, making the BP-optimal information structure achieve optimality in the long run. The key reason is that under the BP-optimal structure, no outcome leads the Receiver to find the alternative model more likely.

We then revisit the seller-buyer example introduced above, where the Sender's preferred action is revealing. Unlike in the previous case, this example illustrates that the BP-optimal structure entails a risk of Receivers switching, which leads them to choose the non-revealing action. Once this occurs, Receivers permanently adopt the alternative due to the identical marginal distributions of signals.

\textbf{Proposition~\ref{prop:seller}} identifies a condition under which the BP-optimal information structure is outperformed by full disclosure and is not optimal in the long run. While full disclosure yields a lower period expected utility for the Sender than the BP-optimal structure, it completely prevents switching(\textbf{Lemma~\ref{lem:extreme}}). When the switching threshold is sufficiently low, Receivers are highly likely to switch, prompting the patient Sender to prefer full disclosure over the BP-optimal structure.


\indent The rest of the paper is structured as follows: Section II introduces our basic model setup and formalizes the Sender's problem. Section III consists of the analysis of the long-run persuasion problem. Section IV concludes with a discussion. All proofs are provided in Appendix~\ref{sec:AppendixProofs}.

\subsection*{Relevant Literature}\
\indent This paper contributes to the strand of Bayesian persuasion research by incorporating dynamic settings and model misspecification. \cite{kamenica2011bayesian} introduced the BP model, which has inspired a vast literature.\footnote{See \cite{kamenica2019bayesian} for a general review of the BP literature.}

\cite{ba2025strategic} is the closest to our work, as they also adopt the Bayes factor rule from \cite{Ba2023} to capture the Receiver's model switching. In their model, the Receiver is uncertain whether the Sender is biased and compares two competing models: selective disclosure (biased and correct) and full disclosure (unbiased and incorrect). The biased Sender aims to induce the Receiver to adopt the misspecified model in the long run, which is more favorable to him.
In contrast, in our model, the Sender prefers to keep the Receiver adopting the correct model, as the alternative is entirely uninformative and strictly worse for the Sender.

\indent Several papers study dynamic persuasion with evolving states \citep{Ely2017Beeps, renault2017optimal, ely2020moving, orlov2020persuading, smolin2021dynamic, ball2023dynamic}, or sequential information disclosure with stationary states \citep{au2015dynamic, honryo2018dynamic, bizzotto2021dynamic}.

More closely related to our model are papers where the state realization is i.i.d. over time, and Receivers observe past outcomes in the repeated cheap talk framework \citep{meng2021value, best2024persuasion, mathevet2022reputation}. \cite{meng2021value} considers long-lived Sender and Receiver. \cite{best2024persuasion} show that the BP outcome can be achieved through coarse feedback about past communication in the absence of commitment. \cite{mathevet2022reputation} examine how reputation management can substitute for commitment when Receivers update the beliefs about the Sender's type.
In contrast, our model features a committed Sender to a fixed information structure, but Receivers doubt it and may switch to a misspecified alternative.

\indent Our paper is also related to the literature on model selection and paradigm shifts. Several papers study how decision-makers switch between decision-making models and how misspecified models can persist \citep{ortoleva2012modeling, cho2015learning, montiel2022competing, Ba2023}. We adopt the Bayes factor rule from \cite{Ba2023} to capture how Receivers switch between competing information structures.

A different strand of literature explores how the Sender influences the Receiver’s model or interpretation. \cite{galperti2019persuasion} examines how the Sender can shape the Receiver's prior beliefs. \cite{schwartzstein2021using} and \cite{aina2023tailored} study how the Sender affects the Receiver’s interpretation of exogenously generated signals. \cite{ichihashi2021design} considers the Sender's design of both signals and suggested interpretations. Unlike these papers, the Sender in our model cannot affect how Receivers interpret signals. Instead, we examine how the Sender optimally designs an information structure and endogenously generates signals, while anticipating the possibility that Receivers may switch between two competing models.

\section{\textbf{Model}}\label{sec:themodel}

\subsection{Setup}\

\indent We consider a long-lived Sender (``he''), who discounts future payoffs at a rate $0 <\delta < 1 $, interacting with an infinite sequence of short-lived Receivers (each ``she''). In each period $t \in \{1,2,\dots\}$, the state $\omega_{t}$ is realized independently across periods, drawn from a commonly known distribution $\mu_0$ over the set $\Omega = \{ \omega^{1}, \omega^{2} \}$. For simplicity, we denote $\mu = \mu (\omega^1)$.

The Receiver in period $t$ chooses an action $a_t$ from the action set $A = \{ a^{1}, a^{2} \}$. The Sender's and Receiver's period utilities at period $t$ are given by $v_t = v(\omega_t, a_t)$ and $u_t = u(\omega_t, a_t)$, respectively. We assume that their utility functions do not change over periods. 

\indent The Sender aims to influence the Receiver's action choice by designing an information structure. An information structure consists of a finite signal space $S$ and a mapping $P: \Omega \rightarrow \Delta (S) $, which specifies a family of distributions $\{ p(\cdot|\omega) \}_{\omega \in \Omega}$ over $S$. For notational convenience, we denote the mapping $P$ as a matrix, i.e., $P = \big( p(\omega, s) \big)_{|\Omega|\times |S|} $ where $p(\omega, s) = Pr(s|\omega)$.

\indent At $t=0$, the Sender chooses an information structure $(S, P)$ and publicly announces it to Receivers. Once chosen, the Sender cannot deviate from this structure. That is, in each period $t=1, 2, \dots$, signal $s_t$ is generated independently according to $P$.\footnote{One practical justification for this assumption is the high fixed cost associated with setting up or changing an information structure. For instance, consider a used car seller (Sender) attempting to persuade the buyers (Receiver) to purchase his cars. He needs to equip an inspection system to send signals regarding the quality of the vehicles. The high fixed cost of implementing such a system may make it impractical to change information structures over time.}


\indent In each period $t$, the Receiver observes the signal realization $s_t$ and updates her belief $\mu_{s_t}$ about the true state at that period using Bayes' rule. Each Receiver is assumed to be a myopic decision maker: She chooses her action to maximize her period utility.
Hence, the Receiver's optimal action after observing signal $s_t$ is given by:
\begin{equation*}
    a_{t}^{*}(\mu_{s_t}) \in \arg\max_{a\in A} \bigg[\mu_{s_t} u(\omega^1 , a) + \big(1- \mu_{s_t} \big) u(\omega^2, a)\bigg].
\end{equation*}
Following \cite{kamenica2011bayesian}, we assume that if the Receiver is indifferent between $a^{1}$ and $a^{2}$, she chooses an action preferred by the Sender.\\

\noindent\textbf{Alternative Information Structure.} \ Departing from the standard BP model, we explore the long-run interaction between the Sender and Receivers, where Receivers do not fully trust the Sender's announced structure.  Specifically, Receivers suspect that signals may be uninformative. Thus, Receivers consider two competing information structures: the Sender's announced structure $(S, P)$ and an \textbf{alternative structure} $(S, \hat{P})$. $(S, \hat{P})$ has the same set of signals as $(S, P)$, but $\hat{P}$ is uninformative while inducing the same signal distributions as $P$. Since the two information structures share the same signal space, we will use $P$ and $\hat{P}$ instead of $(S, P)$ and $(\hat{S}, \hat{P})$ when comparing two information structures hereafter.

Receivers initially adopt $P$ and update their beliefs accordingly. However, they continually compare between $P$ and $\hat{P}$, switching between the two based on past outcomes. We refer to the information structure the Receiver currently uses for belief updating as her \textbf{perceived information structure}.\\

\noindent \textbf{Example.} Assume $\mu_0 = 0.3$. Consider the following information structure as the Sender's choice $(S, P)$: $S = \{s^1, s^2, s^3 \} $, and
\begin{displaymath}
    P= \ \bordermatrix{
    & \ s^1     & s^2   & s^3     \cr
    \omega^1    & 1/3    & 2/3     & 0     \cr
    \omega^2    & 1/7    & 2/7     & 4/7    \cr
            }.
\end{displaymath}
Under this information structure, signals $s^1$, $s^2$, and $s^3$ are realized with probabilities of 0.2, 0.4, and 0.4, respectively. Then, the alternative information structure $(S,\hat{P})$ is represented as:
\begin{displaymath}
    \hat{P}= \ \bordermatrix{
    & \ s^1     & s^2   & s^3     \cr
    \omega^1    & 0.2    & 0.4  & 0.4     \cr
    \omega^2    & 0.2    & 0.4  & 0.4    \cr
            }.
\end{displaymath}
Note that the Receiver's posterior will always be equal to the prior under $\hat{P}$.

\subsection{Observation and History} \
\indent Since the state and signal realizations are i.i.d. across periods, each Receiver updates their posteriors independently unless they switch their perceived information structure. The only channel through which previous states and signals affect future posteriors is by shaping the Receiver's perceived information structure. Specifically, the Receiver observes the previous outcomes and determines the perceived information structure based on them.

\begin{definition}
    An \textbf{observation at period $t$} is a triplet $o_t \equiv (s_t, a_t, u_t)$ consisting of the realized signal $s_t$, the Receiver's action $a_t$, and the Receiver's utility $u_t$ at period $t$. \textbf{A history at period $t$} is $h_t \equiv \{ o_{\tau} \}_{\tau=1}^{t-1} = (s_{\tau}, a_{\tau}, u_{\tau})_{\tau=1}^{t-1}$, a collection of the observations from period 1 to $t-1$.
\end{definition}

\indent Note that the empirical distribution of state-signal pairs is the crucial clue for inferring the true information structure. However, $o_t$ only involves $s_t$ but not $\omega_t$, implying that the Receiver can directly observe the realized signal but not the true state of the world in each period. Instead, she can infer the true state through her utility at that period, which depends on her chosen action. If an action yields different period utilities across the states, she can deduce the realized state. Otherwise, she only knows the realized signal.

\begin{definition} An action $a \in A$ is \textbf{revealing} if $u(\omega^{1},a) \neq u(\omega^{2},a)$. Otherwise, $a$ is \textbf{non-revealing}.
\end{definition}

\indent Depending on the associated action, each observation $o_t$ contains a different amount of information relevant to the true signal-generating process. To emphasize this, we simplify $o_t$ based on its $a_t$. If $a_t$ is revealing, then $o_t$ is reduced to $(s_t, \omega_t)$. In contrast, if $a_t$ is non-revealing, then $o_t$ is simplified to $s_t$.

\indent Let $(S, Q)$ be an information structure. The likelihood of an observation $o_{t}$ under $(S, Q)$ when the associated action $a_t$ is revealing is given by:
\begin{equation}
    l(o_{t}|S, Q) = \mu_{0} (\omega_{t}) q(\omega_{t}, s_{t}).
\end{equation}
On the other hand, if the associated action $a_{t}$ is non-revealing, the likelihood of the observation $o_{t}$ under $(S, Q)$ is:
\begin{equation}
    l(o_{t}|S, Q) = \mu_{0} q(\omega^1, s_{t}) + (1-\mu_{0}) q(\omega^2, s_{t}).
\end{equation}
From the independence of the states and signals across the periods, the likelihood of history $h_t$ under $(S, Q)$ is the product of the likelihoods of the previous observations:
\begin{equation}
    l(h_t|S, Q) = \prod_{{\tau}=1}^{t-1} l(o_{\tau}|S, Q).
\end{equation}

\subsection{The Receiver's switching rule} \
\indent We employ the Bayes factor rule from \cite{Ba2023} to model the Receiver's switching behavior. At the beginning of each period $t$, before receiving a new signal, the Receiver observes the history at period $t$. Then, she calculates the likelihood of the history under each information structure $P$ and $\hat{P}$ to derive the Bayes factor, which is used to determine the perceived information structure. The Bayes factor, given history $h_t$, is defined as:
\begin{equation}
    \lambda (h_t) = \frac{l(h_t|S, \hat{P})}{l(h_t|S, P)}.
\end{equation}

The Bayes factor $\lambda(h_t)$ measures the relative likelihood of history under $\hat{P}$ compared to $P$. The higher value of $\lambda(h_t)$ indicates that the history $h_t$ is more likely under $\hat{P}$ than under $P$.\\


\noindent\textbf{Switching Rule.} \ Let $\alpha \geq1$ be a threshold for switching. Suppose that at period $\hat{t}-1$, the perceived information structure is $P$ and $\lambda(h^{\hat{t}}) > \alpha $ at $\hat{t}$. Then, the Receiver switches her perceived information structure from $P$ to $\hat{P}$ at $\hat{t}$ and updates her beliefs based on $\hat{P}$. Once the shift occurs, $P$ becomes the alternative information structure, and Receivers will compare $\hat{P}$ with $P$. They will shift back to $P$ if $\hat{\lambda} (h_t) \equiv \frac{1}{\lambda(h_{t})} = \frac{l(h_t|P)}{l(h_t|\hat{P})} > \alpha$. Importantly, some histories may lead the Receiver to a misperceived information structure, even if she starts with the correct one.

\begin{remark}
    If the Receiver chooses a non-revealing action at period $t$, she only knows the realized signal at period $t$. Since both information structures induce the same signal distributions, she learns nothing about the true structure of that period, meaning that $\lambda(h_t) = \lambda(h_{t-1})$.
\end{remark}

\subsection{Persistence of Information Structures}\

\indent Let $H_t$ be the set of all possible histories at period $t$, and let $A_t^{\alpha}(P) \subseteq H_t$ be the subset of histories in which the perceived information structure at $t$ is $P$. Then, $A_t^{\alpha}(\hat{P})=\big({A_t^{\alpha}(P)}\big)^{c}$ corresponds to the histories where the perceived information structure at $t$ is $\hat{P}$. The probability of the Receiver adopting $P$ at $t$ is defined as:
\begin{equation*}
    l\left(A_t^{\alpha}(P)\right) \equiv \int_{h_t \in A_t^{\alpha}(P)} l(h_t|P) \nu(dh_t),
\end{equation*}
where $\nu$ is a probability measure over $H_t$. Note that $l(A_1(P)) = 1$. \ footnote {Note that $P$ is the data-generating information structure. Therefore, the probability of a history $h_t$ occurring is $l(h_t|P)$.}

\indent Under the Receiver's switching rule, her perceived information can shift between $P$ and $\hat{P}$ multiple times, depending on history. However, some information structures can be robust enough to prevent switching regardless of history. Also, it is possible for the Receiver to eventually settle on one information structure after several shifts. These concepts are especially important to analyze the Sender's strategy when he is patient($\delta \rightarrow 1$).

\begin{definition}
Let $P$ be an information structure and $\alpha>1$ a switching threshold. 
\begin{enumerate}[label={(\arabic*)}]
    \item $P$ \textbf{persists} if $l(A_t^{\alpha}(P)) = 1$ for any $t = 1, 2, \dots$.
    \item $P$ \textbf{eventually persists} if $\lim_{t\rightarrow\infty}l(A_t^{\alpha}(P)) = 1$.\footnote{Our definition of persistence is stricter than that in \cite{Ba2023}. In \cite{Ba2023}, the persistence of a model requires a positive probability that the decision-maker sticks to the model in the long run. In contrast, our definition requires that the Receiver almost surely adheres to the information structure in the long run.} 
\end{enumerate}
\end{definition}


\subsection{Sender's Problem} \
\indent The Sender strategically chooses the information structure to maximize his lifetime expected utility, accounting for the risk of the Receiver switching her perceived information structure. To simplify the formulation of the Sender's problem, we introduce the following notations.


\indent The Sender's expected utility when both the Sender and Receiver share the same belief $\mu$ is defined as:
\begin{displaymath}
    \hat{v}(\mu) \equiv \mu v\big(\omega^1, a^{*}(\mu) \big) + \big(1 - \mu \big) v\big(\omega^2, a^{*}(\mu)\big).
\end{displaymath}
Let $\pi_p \in \Delta\left( \Delta(\Omega) \right)$ denote the distribution over posteriors induced by $P$ and given the prior $\mu_0$. The Sender's period expected utility, conditional on the perceived information structure $(S, P)$ and prior $\mu_0$, is defined as:
\begin{displaymath}
    V(\mu_0 |S, P) \equiv \sum_{\mu \in Supp(\pi_{P})} \pi_P (\mu) \hat{v}\big(\mu \big).
\end{displaymath}
Similarly, let $V(\mu_0 | S, \hat{P})$ denote the Sender’s period expected utility when the perceived information structure is $(S, \hat{P})$.

Finally, we define the Sender's problem as:
\begin{equation}
\begin{split}
    \max_{(S,P)} & \ \tilde{V}_{\alpha,\delta}(\mu_0|S,P) \equiv \sum_{t=1}^{\infty} \delta^{t-1} \bigg[ l(A_t^{\alpha}(P)) \ V(\mu_0 |S, P) + \big( 1-l(A_t^{\alpha}(P)) \big) \ V(\mu_0 |S, \hat{P}) \bigg] \\
    \text{s.t.}& \quad (1)\ \sum_{\mu\in Supp(\pi_P)} \mu \cdot \pi_P(\mu) = \mu_0.
\end{split}
\end{equation}
Here, the constraint ensures the Bayes plausibility of the distribution $\pi_P$. Since $P$ is the true information structure, $\pi_P$ corresponds to the actual distribution over posteriors.

\section{\textbf{Analyses}}\label{sec:analyses 1}
\subsection{Extreme information structures}\
\indent We begin our analysis by examining the long-run optimality of the information structure that solves the one-shot BP problem (BP-optimal). One sufficient condition for its optimality is persistence. If the Receiver adheres to the BP-optimal information structure without ever switching, the Sender can maximize his period expected utility in every period by choosing the BP-optimal policy. Consequently, selecting such an information structure maximizes the Sender's lifetime expected utility. The following lemma establishes that extreme information structures, namely full disclosure and no disclosure, persist.

\begin{lemma}\label{lem:extreme}
    If $P$ is either full disclosure or no disclosure, then it persists.
\end{lemma}

\indent Since the alternative of no disclosure is identical to itself, it obviously persists. Additionally, under full disclosure, every possible observation is always more likely to be generated by the true information structure than by the alternative. As a result, $\lambda(h_t)$ weakly decreases over time, preventing the Receiver from switching. This lemma demonstrates the existence of persistent information structures and shows that the sum of discounted period expected utilities from these extreme structures serves as a lower bound for the Sender's lifetime expected utility. An immediate optimality result follows from \textbf{Lemma~\ref{lem:extreme}}.

\begin{corollary} \label{cor:extreme}
    If the BP-optimal information structure is full disclosure or no disclosure, then it is optimal in the long run.
\end{corollary}



\cite{kamenica2011bayesian} provide the conditions under which extreme information structures are BP-optimal. If $\hat{v}$ is concave or takes the same value as its concave closure at the prior $\mu_0$, no disclosure is BP-optimal.\footnote{The concave closure of $\hat{v}$ is $V(\mu) \equiv \sup \{ z|(\mu,z) \in co(\hat{v}) \}$, where $co(\hat{v}$) denotes the convex hull of the graph of $\hat{v}$. This optimality criterion can be expressed as $\hat{v}(\mu_0) = V(\mu_0)$.} Conversely, if $\hat{v}$ is convex, full disclosure is BP-optimal. Given that both policies persist, they remain optimal in the long run. Note that when no disclosure is BP-optimal, the Sender cannot benefit from persuasion in the long run. In contrast, when $\hat{v}$ is strictly convex, the Sender can benefit from persuasion by utilizing full disclosure.


\subsection{When every action is revealing}\
\indent To analyze the optimality of non-extreme BP-optimal information structures, it is crucial to consider what the Receiver can observe in each period, which depends on her actions. The Receiver's action set can be classified into two cases: (i) both actions are revealing, or (ii) only one action is revealing.\footnote{If every action is non-revealing, the state does not affect the Receiver's utility, and she does not care about the state. Thus, we assume that at least one action is revealing.} We first examine the case where every action is revealing. 

\begin{lemma} \label{lem:revealing}
    If every action is revealing, the chosen information structure $P$ eventually persists.
\end{lemma}

\indent Recall that the Bayes factor $\lambda(h_t)$ is a likelihood ratio, where the denominator corresponds to the true data-generating process. This implies that although the Receiver may adopt the alternative information structure in early periods, $\lambda(h_t)$ almost surely converges to zero. Consequently, Receivers eventually adhere to $P$. \textbf{Lemma~\ref{lem:revealing}} implies that the BP-optimal information structure can outperform any other information structure if the Sender is sufficiently patient.

\begin{proposition} \label{prop:revealing}
    Suppose that every action is revealing, and let $(S, P)$ be a BP-optimal information structure. Then, for any non-BP-optimal $(S', P')$, there exists $\delta'\in (0,1)$ such that for all $\delta\geq \delta'$,
    \begin{equation*}
        \tilde{V}_{\alpha,\delta}(\mu_0| S, P) \geq \tilde{V}_{\alpha,\delta}(\mu_0|S', P').    
    \end{equation*}
    
\end{proposition}

\begin{remark}
    The results established in \textbf{Lemma~\ref{lem:extreme}, Corollary~\ref{cor:extreme}, Lemma~\ref{lem:revealing}, and Proposition~\ref{prop:revealing}} extend to environments with more than two states and actions. The key arguments, relying on persistence and the convergence of the Bayes factor, remain valid under a general finite-state, finite-action framework.
\end{remark}


\subsection{When the Sender prefers the non-revealing action}\
\indent We now examine cases in which one action is non-revealing and, furthermore, preferred by the Sender.

\subsubsection{Speed Limit Enforcement Example}\
\indent Consider an infinite interaction between a Traffic Authority (Sender) and drivers (Receiver). The state of the world is either enforcement ($E$) or no enforcement ($NE$). The driver's action set is $A = \{S, NS \}$, where $S$ denotes speeding, and $NS$ denotes not speeding. The prior belief is $\mu_0(E) = 0.3$. Table~\ref{tab:speed} presents the period utilities of both players under different states and actions.

\begin{table}[!htb]
    \begin{minipage}{.5\linewidth}
      \centering
        \begin{tabular}{  c | c  c  }
            \thickhline
             & \quad S \quad & \quad NS \quad\\
             \thickhline
            \quad E  \quad & \quad 0 \quad  & \quad 1 \quad \\
            \quad NE \quad & \quad 0 \quad  & \quad 1 \quad \\
             \thickhline
        \end{tabular}\\
        \caption*{Traffic Authority}
    \end{minipage}%
       \begin{minipage}{.5\linewidth}
        \centering
        \begin{tabular}{  c | c  c  }
            \thickhline
             & \quad S \quad & \quad NS \quad\\
             \thickhline
             \quad E \quad  & \quad -1 \quad  & \quad 0 \quad \\
             \quad NE \quad & \quad 1  \quad  & \quad 0 \quad \\
             \thickhline
        \end{tabular}\\
        \caption*{Driver}
    \end{minipage}%
    \caption{Period Utilities of the Speed Limit Enforcement Example}
    \label{tab:speed}
\end{table}

\indent In each period, the driver chooses between $S$ and $NS$. Note that $S$ is revealing, while $NS$ is non-revealing. The Traffic Authority strictly prefers $NS$ to $S$ in both states. Under the prior $\mu_0$, the driver chooses to speed ($a^{*} (\mu_0) = S$).

\indent The BP-optimal information structure $(S, P)$ uses two signals -- sign ($s$) and no-sign ($ns$)-- with the following conditional probabilities:
\begin{displaymath}
     P= \ \bordermatrix{
    & s    & ns    \cr
    E    & 1     & 0     \cr
    NE   & 3/7     & 4/7    \cr
            }.
\end{displaymath}
The corresponding alternative $\hat{P}$ is:
\begin{displaymath}
     \hat{P}= \ \bordermatrix{
    & s    & ns    \cr
    E    & 0.6     & 0.4    \cr
    NE   & 0.6     & 0.4    \cr
            }.
\end{displaymath}

Initially, the driver's perceived information structure is $P$. Upon observing $s$, her posterior is $\mu_{s}(E) = 0.5$, and her optimal action is $a^{*}(\mu_{s}) = NS$, which is non-revealing. If she observes $ns$, the posterior is $\mu_{ns}(E) = 0$, and she chooses $a^{*}(\mu_{ns}) = S$, which is revealing. The driver, therefore, anticipates three types of observations in each period: $s$, $(ns, E)$, or $(ns, NE)$. Importantly, since $P$ is the true signal-generating structure, $(ns, E)$ never occurs. Table~\ref{tab:speedlikeli} presents the likelihoods of observations under $P$ and $\hat{P}$.

\begin{table}[!ht]
\centering
\begin{threeparttable}
\begin{tabular}{  C{7cm} | C{2.5cm}  C{2.5cm}  } 
  \hline
  Observation & $P$ & $\hat{P}$ \\ 
  \hline
  $s$ & 0.6 & 0.6 \\ 
  $(ns, NE)$ & 0.4 & 0.28 \\
  $(ns, E)$ & 0 & 0.12\\
  \hline
\end{tabular}
\begin{tablenotes}
    \item \footnotesize{\textit{Notes}: $P$ is the optimal information structure under BP, and $\hat{P}$ is a no-information structure generating the same signal distribution as $P$. The observations are what the driver anticipates when her actions are based on updating her beliefs according to $P$.}
\end{tablenotes}
\end{threeparttable}
\caption{Likelihoods of Observations under Announced and Alternative Information Structures in the Speed Limit Enforcement Example}
\label{tab:speedlikeli}
\end{table}

Comparing the likelihoods of possible observations, we observe that $l(o|P) \geq l(o|\hat{P})$ for all possible observation $o$. This holds because $(ns, E)$ is more likely to occur under $\hat{P}$, but it never occurs under the true information structure $P$. Therefore, for any history $h_t$, the Bayes factor $\lambda(h_t)$ never exceeds one, nor does it reach the threshold $\alpha$. This implies that $P$ persists, leading to it being optimal in the long run.

\subsubsection{Generalizing the Speed Limit Enforcement Example}\
\indent The following proposition generalizes the result from the speed limit enforcement example.

\begin{proposition} \label{prop:speed}
    Suppose that there exists an action $a$ such that:
    \begin{enumerate}[label=(\roman*)]
        \item $a$ is non-revealing, and
        \item $v(\omega, a) > v(\omega, \hat{a})$ for all $\omega \in \Omega$ and $a\neq \hat{a}$. 
    \end{enumerate}
   Then, the BP-optimal information structure is optimal $\forall \delta\in(0,1)$.
\end{proposition}

\begin{remark}
    Even though our environment assumes that the alternative information structure is totally uninformative, \textbf{Proposition~\ref{prop:speed}} extends to any alternative that generates the same marginal distributions over signals as $P$. Whenever the alternative structure induces the same signal distributions, the BP-optimal information structure persists and is thus optimal in the long run. The proof is similar.
\end{remark}

\subsection{When the Sender prefers the revealing action}\
\indent Now, we revisit the seller-buyer example from the introduction. In this example, the BP-optimal information structure lacks persistence, challenging its long-run optimality. In contrast to the speed limit enforcement example, the Sender now prefers the revealing action over the non-revealing one. In this case, selecting the BP-optimal information structure poses the risk of the Receiver switching. Moreover, as illustrated below, once the Receiver switches from the Sender's announced information structure, she never reverts to it.

\subsubsection{Seller-Buyer Example} \
\indent A seller (Sender) and a buyer (Receiver) interact infinitely many times. The state of the world represents whether the quality of the product is high ($H$) or low ($L$). The buyer chooses between buying ($B$) and not buying ($NB$) in each period. The prior belief is $\mu_0(H) = 0.3$. Table~\ref{tab:sellerutil} shows the period utilities for both players. Note that the seller always prefers the buyer to choose $B$, which is revealing. In contrast, the seller's less favored action, $NB$, is non-revealing.

\begin{table}[!ht]
    \begin{minipage}{.5\linewidth}
      \centering
        \begin{tabular}{  c | c  c  }
            \thickhline
             & \quad B \quad & \quad NB \quad\\
             \thickhline
            \quad H  \quad & \quad 1 \quad  & \quad 0 \quad \\
            \quad L \quad & \quad 1 \quad  & \quad 0 \quad \\
             \thickhline
        \end{tabular}\\
        \caption*{Seller}
    \end{minipage}%
       \begin{minipage}{.5\linewidth}
        \centering
        \begin{tabular}{  c | c  c  }
            \thickhline
             & \quad B \quad & \quad NB \quad\\
             \thickhline
            \quad H  \quad & \quad 1 \quad  & \quad 0 \quad \\
            \quad L \quad & \quad -1 \quad  & \quad 0 \quad \\
             \thickhline
        \end{tabular}\\
        \caption*{Buyer}
    \end{minipage}%
    \caption{Period Utilities of the Seller-Buyer Example}
    \label{tab:sellerutil}
\end{table}

\subsubsection{BP-optimal information structure} \
\indent The BP-optimal information structure consists of signal space $S = \{h,l \}$ and the following conditional probabilities $P$:
\begin{displaymath}
    P= \ \bordermatrix{
    & h     & l    \cr
    H    & 1     & 0     \cr
    L     & 3/7    & 4/7    \cr
            }.
\end{displaymath}
If the perceived information structure is $P$, the buyer purchases the product upon receiving $h$ ($a^{*}(\mu_{h}) = B$). Conversely, observing $l$ reveals with certainty that the product quality is low, so the buyer does not purchase after observing $l$ ($a^{*}(\mu_{l}) = NB$). Given that the probability of receiving $h$ is 0.6, the seller's period expected utility is 0.6.

\indent The corresponding alternative information structure $\hat{P}$ is:
\begin{displaymath}
    \hat{P}= \ \bordermatrix{
    & h     & l    \cr
    H    & 0.6     & 0.4     \cr
    L     & 0.6    & 0.4    \cr
            }.
\end{displaymath}
If the perceived information structure is $\hat{P}$, the buyer never buys the product, regardless of the signal, which leads to the seller's period expected utility being 0. Therefore, the seller's lifetime expected utility from choosing $P$ is
\begin{displaymath}
        \begin{split}
            \tilde{V}_{\alpha,\delta}(\mu_0|S, P) &= 0.6 + 0.6 \delta l(A_2(P)) + 0.6 \delta^2 l(A_3(P)) + \cdots\\
            &= 0.6 \sum_{t=1}^{\infty} \delta^{t-1} l(A_{t}(P)).
        \end{split}
\end{displaymath}

The buyer expects the possible observations of $(h, H)$, $(h, L)$, or $l$ in each period when her perceived information structure is $P$. Table~\ref{tab:sellerlikelihood} presents the likelihoods of each observation under $P$ and $\hat{P}$. Note that every observation is feasible under the true $P$. Observation $(h, H)$ is more likely to be observed under $P$, and $(h, L)$ is more likely to be observed under $\hat{P}$. Thus, $(h, H)$ decreases the Bayes factor $\lambda(h_t)$ while $(h, L)$ increases $\lambda(h_t)$.

\begin{table}[!ht]
\centering
\begin{threeparttable}
\begin{tabular}{  C{7cm} | C{2.5cm}  C{2.5cm}  } 
  \hline
  Observation & $P$ & $\hat{P}$ \\ 
  \hline
  $(h,H)$ & 0.3 & 0.18 \\ 
  $(h,L)$ & 0.3 & 0.42 \\
  $l$     & 0.4 & 0.4 \\ 
  \hline
\end{tabular}
\begin{tablenotes}       
    \item \footnotesize{\textit{Notes}: $P$ is the optimal information structure under BP, and $\hat{P}$ is a no-information structure generating the same signal distribution as $P$. The observations are what the driver anticipates when her actions are based on updating her beliefs according to $P$.}
\end{tablenotes}
\end{threeparttable}
\caption{Likelihoods of Observations under Announced and Alternative Information Structures in the Seller-Buyer Example}
\label{tab:sellerlikelihood}
\end{table}

If a history contains a substantially greater number of $(h, L)$ than $(h, H)$, the buyer will switch her perceived information structure from $P$ to $\hat{P}$. Notably, once the buyer switches from $P$ to $\hat{P}$, she never switches back. When the perceived information structure is $\hat{P}$, the buyer always chooses the non-revealing action $NB$, so she only observes $h$ or $l$ each period. Since the signal distributions under $P$ and $\hat{P}$ are identical, $\hat{\lambda}(h_t)$ remains constant, preventing a return to $P$. This implies that the seller faces a risk of getting zero forever once the buyer switches away from $P$.

\subsubsection{Comparing BP-optimal with Full Disclosure}\
\indent Instead of the BP-optimal information structure, consider full disclosure. The true information structure $P_{f}$ and its alternative $\hat{P}_{f}$ are:
\begin{displaymath}
     P_{f}= \ \bordermatrix{
    & h    & l    \cr
    H    & 1     & 0     \cr
    L   & 0     & 1    \cr
            }, \quad \quad
    \hat{P}_{f}= \ \bordermatrix{
    & h     & l    \cr
    H    & 0.3     & 0.7     \cr
    L   & 0.3     & 0.7    \cr
            }.
\end{displaymath}
If the perceived information structure is $P_{f}$, the buyer purchases only when the signal indicates high quality, resulting in a period expected utility of 0.3 for the seller. Since $P_{f}$ consists (\textbf{Lemma~\ref{lem:extreme}}), the seller's lifetime expected utility using full disclosure is:
\begin{displaymath}
    \tilde{V}_{\alpha,\delta}(\mu_0|S,P_f) =  \sum_{t=1}^{\infty} 0.3\delta^{t-1} = \frac{0.3}{1-\delta}.
\end{displaymath}

\indent In this environment, $A_t^{\alpha}(P) = \{ h_t \in H_t \ | \ \lambda(h_s) < \alpha \ \forall s\leq t\}$, and $l(A_t^{\alpha}(P))$ represents the probability that no switching (from $P$ to $\hat{P}$) has occurred until period $t$, since the buyer never returns to $P$ once she switches. It follows that $l(A_t^{\alpha}(P))$ weakly decreases over time. If there exists $T>0$ such that $l(A_T(P))<0.5$, full disclosure $P_f$ yields higher period expected utility than the BP-optimal $P$ for any period $t\geq T$. In such a case, a sufficiently patient seller prefers $P_f$ to $P$, and $P$ fails to achieve the long-run optimality.

Note that $l(A_t^{\alpha}(P))$ weakly increases with the switching threshold $\alpha$ and converges to 1 as $\alpha \rightarrow \infty$ for all $t\geq 1$. In contrast, as $\alpha$ approaches 1, $l(A_t^{\alpha}(P))$ decreases, making it more likely that $P_f$ outperforms $P$. Furthermore, if there exists a switching threshold $\hat{\alpha}$ such that $P_f$ outperforms $P$ under $\hat{\alpha}$, $P$ is not optimal if $\alpha \leq \hat{\alpha}$.

While $l(A_t^{\alpha})(P)$ is crucial for the seller's optimal strategy, direct computation of it is often intractable. To complement the theoretical analysis, Appendix~\ref{sec:AppendixSimul} presents simulation results for the seller-buyer example, providing numerical insights into $l(A_t^\alpha(P))$ and the seller's optimal strategy.

For instance, consider $\alpha<1.4$ in the seller-buyer example. For any history $h_t$ consisting of $o_t = (h,L)$ and $o_s = l$ for all $s<t$, $\lambda(h_t)=1.4>\alpha$, which triggers a permanent switching from $P$ to $\hat{P}$. Note that the likelihoods of such histories are given by $l(h_t|S,P) = 0.4^s \cdot 0.3$. Therefore, $\lim_{t\rightarrow \infty} l(A_t^{\alpha}(P)) < \frac{0.3}{1-0.4} = 0.5$. Consequently, if the seller is patient enough ($\delta\rightarrow 1$), he will never choose $P$ for any $\alpha \leq 1.4$, as he would achieve a higher expected utility with $P_f$.

The following proposition formalizes this result by providing a condition under which the BP-optimal information structure fails to achieve long-run optimality.

\begin{proposition} \label{prop:seller}
    Let $(S, P)$ and $(S_f, P_f)$ be the BP-optimal information structure and full disclosure, respectively. Suppose that:
    \begin{enumerate}[label=(\roman*)]
        \item An action ($a^1$) is revealing, and the other action ($a^2$) is non-revealing,
        \item $a^{*}(\mu_0) = a^2$, and
        \item $v(\omega^1, a^1) = v(\omega^2, a^1) > v(\omega, a^2)$ for all $\omega\in\Omega$.
    \end{enumerate}
    Then, there exists $\hat{\alpha} > 1$ and $\hat{\delta} \in (0,1)$ such that for any switching threshold $\alpha \leq \hat{\alpha}$ and any $\delta > \hat{\delta}$ ,
    \begin{equation*}
        \tilde{V}_{\alpha,\delta}(\mu_0| S_f, P_f) > \tilde{V}_{\alpha,\delta}(\mu_0| S, P).
    \end{equation*}
\end{proposition}

Choosing between the BP-optimal $P$ and full disclosure $P_f$ involves a fundamental trade-off. While $P$ offers a higher period expected utility if the Receiver adopts it, it also exposes the Sender to the risk of permanent switching. The condition $v(\omega^1, a^1) = v(\omega^2, a^1)$ ensures that the period utility advantage under $P$ is bounded, which in turn guarantees the existence of such a switching threshold $\hat{\alpha}$.\footnote{If $v(\omega^1, a^1) \neq v(\omega^2, a^1)$, it is possible that the difference between $V(\mu_0|S, P)$ and $V(\mu_0|S, P_f)$ is significant so that $l(A_t^{\alpha}(P)) \ V(\mu_0 |S, P) + \big( 1-l(A_t^{\alpha}(P)) \big) \ V(\mu_0 |S, \hat{P}) > V(\mu_0|S, P_f)$ for any $\alpha\geq 1$. Then, the Sender prefers $P$ over $P_f$ regardless of the value of $\delta$.}


\section{\textbf{Discussion}}\label{sec:discussion}\
\indent This paper studies a long-run persuasion problem where the Receiver may doubt the Sender's information structure and switch to an uninformative alternative based on observed histories. We show that when the Sender's preferred action is non-revealing, the BP-optimal information structure persists and maximizes the Sender's lifetime expected utility. In contrast, when the preferred action is revealing, the BP-optimal structure does not persist, potentially leading the Sender to prefer full disclosure.

\indent One can consider different alternative information structures. A natural extension is to allow for alternatives that preserve the same signal distribution but are not fully uninformative. It would be worthwhile to examine whether our main results remain valid in such settings. Another direction involves incorporating other sources of uncertainty, such as cheap talk deviations or noisy communication channels. Since these alternatives no longer yield the same signal distributions as the Sender's announcement, the Bayes factor may evolve even after the Receiver takes the non-revealing action, potentially altering the dynamics studied here.

\indent The framework could be extended to allow the Receiver to consider multiple competing alternatives rather than just one. In such cases, her switching behavior could be modeled using maximum likelihood estimation rather than the Bayes factor rule. This extension would facilitate the analysis of how the Sender’s strategy adapts when the Receiver evaluates a broader class of information structures.

\indent Beyond the current framework, an interesting research direction is to consider a forward-looking Receiver who strategically experiments to learn the true information structure. This leads to a multi-armed bandit problem, in which the Receiver chooses actions not only for immediate payoffs but also to gather information. Characterizing equilibrium behavior in such a setting would offer valuable insight into dynamic sender–receiver interactions under endogenous learning.



\bibliographystyle{apalike}
\bibliography{mybibfile}

\newpage

\appendix
\renewcommand{\thesection}{\Alph{section}}
\renewcommand{\thesubsection}{\Alph{section}.\arabic{subsection}:}

\section{Proofs}\label{sec:AppendixProofs}

\subsection{Proof of Lemma\ref{lem:extreme}}
\begin{proof}
    Let $(S, P)$ be the true information structure that the Sender announced and $\mu = Pr(\omega^1)$. First, if $P$ is no disclosure, then $P = \hat{P}$. Therefore, $l(o|P) = l(o|\hat{P})$ for any observation $o$, and $\lambda(h_t) = 1$ for any $t>0$. Thus, it persists.

    Next, suppose that $P$ is a full disclosure policy. Since $P$ is full disclosure, $S$ can be partitioned as $\{ S^1, S^2 \}$, where $S^1 = \{s\in S | \mu_{s}(P) = 1 \} \subset S$ and $S^2 = \{s\in S | \mu_{s}(P) = 0 \} \subset S$. Then, $p(\omega^1, s^2) = 0$ for $s^2 \in S^2$ and $p(\omega^2, s^1) = 0$ for $s^1 \in S^1$. Note that $\hat{p}(\omega^1, s^2) > 0$ for $s^2 \in S^2$ and $\hat{p}(\omega^2, s^1) > 0$ for $s^1 \in S^1$. From $Pr(s|P) = Pr(s|\hat{P})$ for any $s\in S$, we have $p(\omega^i, s^i) > \hat{p}(\omega^i, s^i)$ for $i=1, 2$, where $s^i \in S^i$.
    
    According to the action types, we have two cases. First, assume that every action is revealing, and without loss of generality, assume $u(\omega^i, a^i) > u(\omega^i, a^j)$ for $i\neq j$. Since every action is revealing and $P$ is the true information structure, the Receiver can observe either $(s^1, \omega^1)$ with $s^1 \in S^1$ or $(s^2, \omega^2)$ with $s^2 \in S^2$ when the perceived information structure is $P$. Hence, for any observation $o_{t}$, we either have 
    \begin{displaymath}
    \begin{split}
        &l(o_{t}|P) = \mu_0 p(\omega^1, s^1) > \mu_0 \hat{p} (\omega^1, s^1) = l(o_{t}|\hat{P}), \quad \text{or}\\
        &l(o_{t}|P) = (1-\mu_0) p(\omega^2, s^2) > (1-\mu_0) \hat{p} (\omega^2, s^2) = l(o_{t}|\hat{P}).
    \end{split}
    \end{displaymath}
    Then, for any history $h_t = \{o_{\tau}\}_{\tau=1}^t$, we have
    \begin{equation*}
        \lambda(h_t) = \frac{\prod_{{\tau}=1}^{t-1} l(o_{\tau}|\hat{P})}{\prod_{{\tau}=1}^{t-1} l(o_{\tau}|P)} \leq \frac{\prod_{{\tau}=1}^{t-1} l(o_{\tau}|P)}{\prod_{{\tau}=1}^{t-1} l(o_{\tau}|P)}=1.
    \end{equation*}
    Thus, $Pr\big[ \lambda(h_{t}) \geq \alpha \big] =0$ for any $h_t$ and $t\geq 1$. $P$ persists.

    \indent Next, without loss of generality, assume that action $a^1$ is revealing while $a^2$ is not, and $u(\omega^1, a^1) > u(\omega^1, a^2) = u (\omega^2, a^2) > u (\omega^2, a^1)$. The possible observation in each period is $(s^1, \omega^1)$ or $s^2$ since $a^* (\mu_{s^1}(P)) = a^1$ and $a^* (\mu_{s^2}(P)) = a^2$.
    
    For any observation $o_{t}$, we either have 
    \begin{displaymath}
    \begin{split}
        &l(o_{t}|P) = \mu_0 p(\omega^1, s^1) > \mu_0 \hat{p} (\omega^1, s^1) = l(o_{t}|\hat{P}), \quad \text{or}\\
        &l(o_{t}|P) = Pr(s^2|P) > Pr(s^2|\hat{P}) = l(o_{t}|\hat{P}).
    \end{split}
    \end{displaymath}
    Then, for any history $h_t = \{o_{\tau}\}_{\tau=1}^t$, we have
    \begin{equation*}
        \lambda(h_t) = \frac{\prod_{{\tau}=1}^{t-1} l(o_{\tau}|\hat{P})}{\prod_{{\tau}=1}^{t-1} l(o_{\tau}|P)} \leq \frac{\prod_{{\tau}=1}^{t-1} l(o_{\tau}|P)}{\prod_{{\tau}=1}^{t-1} l(o_{\tau}|P)}=1.
    \end{equation*}
    Thus, $Pr\big[ \lambda(h_{t}) \geq \alpha \big] =0$ for any $h_t$ and $t\geq 1$. $P$ persists.

\end{proof}

\subsection{Proof of Lemma\ref{lem:revealing}}
\begin{proof}
    Let $(S, P)$ be the true information structure that the Sender chose and $(S, \hat{P})$ be the alternative. Note that every observation $o_t$ can be represented as the pair of realized signal and state $(s_t, \omega_t)$. The likelihoods of $o_t$ under $P$ and $\hat{P}$ are $l(o_t|P) = \mu_{0} (\omega_t) p(\omega_t, s_t)$ and $l(o_t|\hat{P}) = \mu_{0} (\omega_t) \hat{p}(\omega_t, s_t)$, respectively. Denote $\mu_0 \equiv \mu_0 (\omega^1)$, $p^i \equiv p(\omega^1, s^i)$, $q^i \equiv p(\omega^2, s^i)$, and $e^i \equiv Pr(s^i)$ for each $s^i \in S$. Then, $\hat{p} (\omega, s^i) = e^i$ for all $\omega \in \Omega$.
    
    We claim that the Bayes factor is a martingale or a supermartingale. Let $\lambda(h_t)$ be the Bayes factor at period $t$. First, suppose $p(\omega, s) \neq 0$ $\forall$ $\omega \in \Omega$, $s\in S$. The expectation of $\lambda(h_{t+1})$ given $h_t$ is
    \begin{displaymath}
        \begin{split}
            \mathbb{E} \bigg[ \lambda(h_{t+1}) \bigg| h_t \bigg] & = \lambda(h_t) \ \mathbb{E} \bigg[ \frac{l(o_{t+1}|\hat{P})}{l(o_{t+1}|P)} \bigg| h_t \bigg]\\
            & = \lambda(h_t) \ \sum_{i=1}^{|S|} \bigg[ \mu_{0} p^i\frac{e^i}{p^i} + (1-\mu_0) q^i\frac{e^i}{q^i} \bigg]\\
            & = \lambda(h_t) \ \sum_{i=1}^{|S|} e^i\\
            & = \lambda(h_t).
        \end{split}
    \end{displaymath}
    Hence, $\lambda(h_t)$ is a martingale.

    Next, suppose that there exists $s\in S$ such that $p(\omega^1, s) = 0 $ or $p(\omega^2, s') = 0$. Let $S^1 \subset S$ be the set of $s \in S$ such that $p(\omega^1, s) = 0$ and $S^2 \subset S$ the set of $s \in S$ such that $p(\omega^2, s) = 0$. The expectation of $\lambda(h_{t+1})$ given $h_t$ is
    \begin{displaymath}
        \begin{split}
            \mathbb{E} \bigg[ \lambda(h_{t+1}) \bigg| h_t \bigg] & = \lambda(h_t) \ \mathbb{E} \bigg[ \frac{l(o_{t+1}|\hat{P})}{l(o_{t+1}|P)} \bigg| h_t \bigg]\\
            & = \lambda(h_t) \bigg[\sum_{i=1}^{|S^1|}(1-\mu_{0})e^i + \sum_{i=1}^{|S^2|}\mu_{0}e^i + \sum_{i=1}^{|S\setminus (S^1 \cup S^2)|} e^i \bigg]\\
            & < \lambda(h_t) \ \sum_{i=1}^{|S|} e^i = \lambda(h_t).
        \end{split}
    \end{displaymath}
    Hence,  $\lambda(h_t)$ is a supermartingale.

    Now, we claim that the Bayes factor $\lambda(h_t)$ converges to 0 as $t\rightarrow \infty$ with probability 1. Since $\lambda(h_t)$ is a nonnegative supermartingale, by the martingale convergence theorem, there exists a random variable $\lambda(h_{\infty}) \in [0, \infty)$ such that $\lambda(h_t) \rightarrow \lambda(h_{\infty})$ as $t\rightarrow \infty$ with probability 1. Taking log of $\lambda(h_t)$, we have
    \begin{displaymath}
        \ln{\lambda(h_t)} = \sum_{i=1}^{t-1} \ln{\bigg[\frac{l(o_i|\hat{P)}}{l(o_i|P)}\bigg]}.
    \end{displaymath}
    
    Since $o_i$'s are independently and identically distributed across periods, $\ln{\bigg[\frac{l(o_i|\hat{P)}}{l(o_i|P)}\bigg]}$ has a common mean, denoted as $m$. By Jensen's inequality with the concavity of the natural logarithm function, we have
    \begin{displaymath}
        m = \mathbb{E} \bigg( \ln{\bigg[\frac{l(o_i|\hat{P)}}{l(o_i|P)}\bigg]} \bigg) < \ln{\bigg(\mathbb{E} \bigg[ \frac{l(o_i|\hat{P)}}{l(o_i|P)} \bigg]  \bigg)}.
    \end{displaymath}
    Note that $\mathbb{E} \bigg[ \frac{l(o_i|\hat{P)}}{l(o_i|P)} \bigg] \leq 1$. Thus, $m <0$. By the strong law of large numbers, $\frac{1}{t}\ln{\lambda(h_t)} \rightarrow m$ as $t \rightarrow \infty$ with probability 1. Then, it follows that $\ln{\lambda(h_t)} \rightarrow -\infty$ as $t\rightarrow \infty$ with probability 1. Since the natural logarithm function is continuous,
    \begin{displaymath}
        \ln{\lambda(h_{\infty})} = \ln \bigg[{\lim_{t\rightarrow \infty} \lambda(h_t)} \bigg] = \lim_{t\rightarrow \infty} \bigg[\ln{\lambda(h_t)} \bigg] = -\infty.
    \end{displaymath}
    Therefore, we conclude that $\lambda(h_{\infty}) = 0$ with probability 1, which means $P$ eventually persists.
\end{proof}

\subsection{Proof of Proposition~\ref{prop:revealing}}
\begin{proof}
    Let $\mu_0 = Pr(\omega^1)$ be a fixed prior, $(S, P)$ be the true and BP-optimal information structure, and $(S, \hat{P})$ be its alternative. First, if $(S, P)$ is either full disclosure or no disclosure, by \textbf{Corollary~\ref{cor:extreme}}, it is optimal in the long run for any $\delta\in(0,1)$.
    
    Let $(S, P)$ be a non-extreme BP-optimal. Assuming every action is revealing, by \textbf{Lemma~\ref{lem:revealing}}, there exists $T>0$ such that $l\left(A_t^{\alpha} (P)\right) = 1$ for any $t\geq T$. Therefore, the Sender's period expected utility is $V(\mu_0 | S, P)$ for any $t \geq T$.

    Consider an information structure $(S', P')$ that is not BP-optimal and its alternative $(S', \hat{P'})$. Note that $V(\mu_0 | S, P) > V(\mu_0 | S', P')$ and $V(\mu_0 | S, P) > V(\mu_0 | S', \hat{P'})$. Therefore, $V(\mu_0 | S, P)$ is an upper bound of the Sender's period expected utility using $(S', P')$.

    Define 
    \begin{displaymath}
    \begin{split}
        & D \equiv V(\mu_0 | S, P) - V(\mu_0 | S', P')\\
        & \tilde{V}_T \equiv \sum_{t=1}^{T-1} \delta^{t-1} \left[ l(A_t^{\alpha}(P)) \ V(\mu_0 |S, P) + \big( 1-l(A_t^{\alpha}(P)) \big) \ V(\mu_0 |S, \hat{P}) \right], \\
        & {\tilde{V}'}_{T} \equiv \sum_{t=1}^{T-1} \delta^{t-1} \left[ l(A_t^{\alpha}(P')) \ V(\mu_0 |S', P') + \big( 1-l(A_t^{\alpha}(P')) \big) \ V(\mu_0 |S', \hat{P'}) \right], \quad \text{and}\\
        & D_T = \tilde{V}_T - {\tilde{V}'}_{T}.
    \end{split}
    \end{displaymath}
    The difference of the lifetime expected utilities between $(S, P)$ and $(S', P')$ satisfies
    \begin{displaymath}
        \begin{split}
            \tilde{V}_{\alpha,\delta}(\mu_0|S,P) - \tilde{V}_{\alpha,\delta}(\mu_0|S',P') & \geq D_T +  \sum_{t=T}^{\infty} \delta^{t-1} D \\
            & = D_T + \frac{\delta^{T-1}}{1-\delta}D.
        \end{split}
    \end{displaymath}
    Note that $\lim_{\delta \rightarrow 1} \frac{\delta^{T-1}}{1-\delta} = \infty$. Since $|D_T| <\infty$ and $D>0$, there exists $\hat{\delta} \in (0,1)$ such that for all $\delta \geq \hat{\delta}$, $\tilde{V}_{\alpha,\delta}(\mu_0|S,P) - \tilde{V}_{\alpha,\delta}(\mu_0|S',P') >0$.
\end{proof}

\subsection{Proof of Proposition~\ref{prop:speed}}
\begin{proof}
    Let $\mu_0 = Pr(\omega^1)$ be a fixed prior, $P$ be the optimal policy under BP, and $\hat{P}$ be the alternative information structure. Suppose $v(\omega, a^1) > v(\omega, a^2) $ for all $\omega \in \Omega$ and $u(\omega^1, a^1) = u(\omega^2, a^1)$.
    
    First, if $a^{*}(\mu_0) = a^1$, no disclosure is optimal under BP. From \textbf{Lemma~\ref{lem:extreme}}, we know that no disclosure $P$ persists. Therefore, this policy can maximize the Sender's period expected utility for every period, which means it is the maximizer for the Sender's lifetime expected utility.

    Next, Suppose that $a^{*}(\mu_0) = a^2$. Without loss of generality, suppose that there exists $\mu>\mu_0$ such that $a^{*}(\mu) = a^1$. Define $\mu^{*} \equiv \inf\{\mu\in [0,1] | \mu>\mu_0 \ \text{and} \ a^{*}(\mu) = a^1 \}$. Note that the Receiver's expected utility is linear in her belief $\mu$. Thus, there exists $\mu'$ such that $\mathbb{E}_{\mu'} \big[u(\omega, a^1) \big] = \mathbb{E}_{\mu'} \big[u(\omega, a^2) \big]$. Moreover, $\mu' = \mu^{*}$, and thus $a^{*}(\mu^{*}) = a^1$.
        
        Using the concavification technique, we derive the BP-optimal information structure $(S, P)$ as
        \begin{displaymath}
            P= \ \bordermatrix{
            & s^1    & s^2    \cr
            \omega^1    & 1     & 0     \cr
            \omega^2   & x     & 1 - x    \cr
                     },
        \end{displaymath}
        where $S =\{s^1, s^2 \}$ and $\frac{\mu_0}{\mu_0 + (1-\mu_0)x} = \mu^*$. Denote $e = \mu_0 + (1-\mu_0)x$. Also, the alternative information structure $\hat{P}$ is
        \begin{displaymath}
            \hat{P}= \ \bordermatrix{
            & s^1    & s^2    \cr
            \omega^1    & e     & 1-e    \cr
            \omega^2   & e      & 1-e    \cr
                     },
        \end{displaymath}
        
        Note that $\mu_{s^1}(P) = \mu^{*}$ and $\mu_{s^2}(P) = 0$. Then, $a^{*} (\mu_{s^1}(P)) = a^1$ and $a^{*}(\mu_{s^2}(P)) = a^2$. Since the true information structure is $P$ and $a^1$ is non-revealing while $a^2$ is revealing, the possible observation is either $s^1$ or $(s^2, \omega^2)$ when perceived information is $P$.
        
        The likelihood of an observation $o = s^1$ is the same between $P$ and $\hat{P}$ since the signal distributions are the same. Also, the likelihoods of $(s^2, \omega^2)$ under $P$ and $\hat{P}$ satisfy
        \begin{displaymath}
        \begin{split}
             l\big( (s^2, \omega^2) |P \big) = (1-\mu_0) (1-x) &= 1 - e \\
             &\geq (1-\mu_0) (1 - e) = l\big( (s^2, \omega^2) |\hat{P} \big).
        \end{split}
        \end{displaymath}
        Therefore, the Bayes factor $\lambda(h_t) \leq 1$ for any history $h_t$. Thus, $P$ persists and is optimal in the long run.
\end{proof}

\subsection{Proof of Proposition~\ref{prop:seller}}
\begin{proof}
    Suppose that $a^1$ is revealing while $a^2$ is non-revealing, $v(\omega^1, a^1) = v(\omega^2, a^1) > \max \{v(\omega^1, a^2),  v(\omega^2, a^2)\}$, and $a^{*}(\mu_0) = a^2$. Also, without loss of generality, suppose that $u(\omega^1, a^1) > u(\omega^1, a^2)$. Denote $\mu = Pr(\omega^1)$, $v^1 = v(\omega^1, a^1) = v(\omega^2, a^1)$, and $v^2 = v(\omega^2, a^2)$.
    
    Let $\mu_0$ be the fixed prior with $\mu_0 \in (0,1)$. Without loss of generality, assume that there exists $\mu >\mu_0$ such that $a^* (\mu) = a^1$. Define $\mu^* \equiv \inf \{\mu \in[0,1] | a^* (\mu) = a^1$\}. Then, the BP-optimal information structure $(S,P)$ is
    \begin{displaymath}
        P= \ \bordermatrix{
    & s^1    & s^2    \cr
    \omega^1    & 1     & 0     \cr
    \omega^2   & x     & 1-x    \cr
            },
    \end{displaymath}
    where $S =\{s^1, s^2 \}$ and $\frac{\mu_0}{\mu_0 + (1-\mu_0)x} = \mu^*$. Denote $e = \mu_0 + (1-\mu_0)x$. Then, the alternative $\hat{P}$ is represented as 
    \begin{displaymath}
        \hat{P}= \ \bordermatrix{
        & s^1    & s^2    \cr
        \omega^1    & e     & 1-e     \cr
        \omega^2   & e     & 1-e    \cr
            }.
    \end{displaymath}


    Denote $l(A_{\infty}(P)) = \lim_{t\rightarrow\infty} l(A_t^{\alpha}(P))$. Note that the Bayes factor $\lambda(h_t)$ evolves in a way that $\lambda(h_{t+1}) = e\lambda(h_t)$ with a probability of $\mu_0$, $\lambda(h_{t+1}) = \frac{e}{x}\lambda(h_t)$ with a probability of $(1-\mu_0)x$, and $\lambda(h_{t+1}) = \lambda(h_t)$ with a probability of $(1-\mu_0)(1-x)$ if $\lambda(h_s)<\alpha$ for all $s\leq t$. Consider a switching threshold $\hat{\alpha} = e/x > 1$. Then, any history $h_t$ such that $o_t = (s^1, \omega^2)$ and $o_s = s^2$ for all $s<t$ leads to the Receiver's permanent switch from $P$ to $\hat{P}$. Thus, the upper bound of $l(A_{\infty}(P))$ is
    \begin{displaymath}
        1 - (1-\mu_0)x [1 + (1-\mu_0)(1-x) + \big((1-\mu_0)(1-x)\big)^2 + \cdots ] = 1 - \frac{(1-\mu_0)x}{e} = \frac{\mu_0}{e}.
    \end{displaymath}
    Take $\epsilon>0$ with $l(A_{\infty}(P)) + \epsilon < \frac{\mu_0}{e}$. Since $l(A_t^{\alpha}(P))$ decreases to $l(A_{\infty}(P))$ and $l(A_{\infty}(P)) < \frac{\mu_0}{e}$, there exists $T>0$ such that for all $t\geq T$, $l(A_t^{\alpha}(P)) < l(A_{\infty}(P)) + \epsilon < \frac{\mu_0}{e}$.

    Define
    \begin{displaymath}
    \begin{split}
        V_t (P) &\equiv l(A_t^{\alpha}(P)) \ V(\mu_0 |S, P) + \big( 1-l(A_t^{\alpha}(P)) \big) \ V(\mu_0 |S, \hat{P})\\
        &= l(A_t^{\alpha}(P)) [ev^1 + (1-e) v^2] + [1-l(A_t^{\alpha}(P))] v^2, \\
        &= l(A_t^{\alpha}(P))e (v^1 - v^2) + v^2,\\
        V_t (P_f) &\equiv \mu_0 v^1 + (1-\mu_0) v^2, \quad \text{and}\\
        D_t &\equiv V_t (P) - V_t (P_f).
    \end{split}
    \end{displaymath}
    Note that for $t\geq T$,
        \begin{displaymath}
        \begin{split}
            D_t &\leq D_T \\
            &=l(A_T(P))e (v^1 - v^2) + v^2 - \big(\mu_0 v^1 + (1-\mu_0) v^2 \big)\\
             &< \frac{\mu_0}{e} e (v^1 - v^2) +  v^2 - \big(\mu_0 v^1 + (1-\mu_0) v^2 \big)\\
             &= 0.
        \end{split}
        \end{displaymath}
    Then, the difference of the lifetime expected utilities between $(S, P)$ and $(S_f, P_f)$ satisfies
    \begin{displaymath}
        \begin{split}
            \tilde{V}_{\alpha,\delta}(\mu_0|S,P) - \tilde{V}_{\alpha,\delta}(\mu_0|S',P') & \leq \sum_{t=1}^{T-1} D_t +  \sum_{t=T}^{\infty} \delta^{t-1} D_T \\
            & = \sum_{t=1}^{T-1} D_t + \frac{\delta^{T-1}}{1-\delta}D_T.
        \end{split}
    \end{displaymath}
    Since $\sum_{t=1}^{T-1} D_t < \infty$, $D_T <0$ and $\lim_{\delta \rightarrow 1} \frac{\delta^{T-1}}{1-\delta} = \infty$, there exists $\hat{\delta}\in (0,1)$ such that for all $\delta\geq\hat{\delta}$, $\tilde{V}_{\alpha,\delta}(\mu_0|S,P) - \tilde{V}_{\alpha,\delta}(\mu_0|S',P') < 0$.
    
    Also, note that $l(A_t^{\alpha}(P))$ (weakly) increases in $\alpha$. ($\because$ For any $h_t \in A_t^{\alpha}(P)$ with $\alpha$, $h_t \in A_t^{\alpha}(P)$ with $\alpha' > \alpha$.)
    Thus, for any switching threshold $\alpha \leq \hat{\alpha}$, $l(A_{T}(P)) < \frac{\mu_0}{e}$, which implies $P_f$ yields higher lifetime expected utility for the Sender than $P$.\\

\end{proof}

\newpage

\section{Simulations}\label{sec:AppendixSimul}
\renewcommand{\thefigure}{B.\arabic{figure}}
\setcounter{figure}{0}
\renewcommand{\thetable}{B.\arabic{table}}
\setcounter{table}{0}

\begin{figure}[htb!]
    \centering
    \includegraphics[width=1\textwidth]{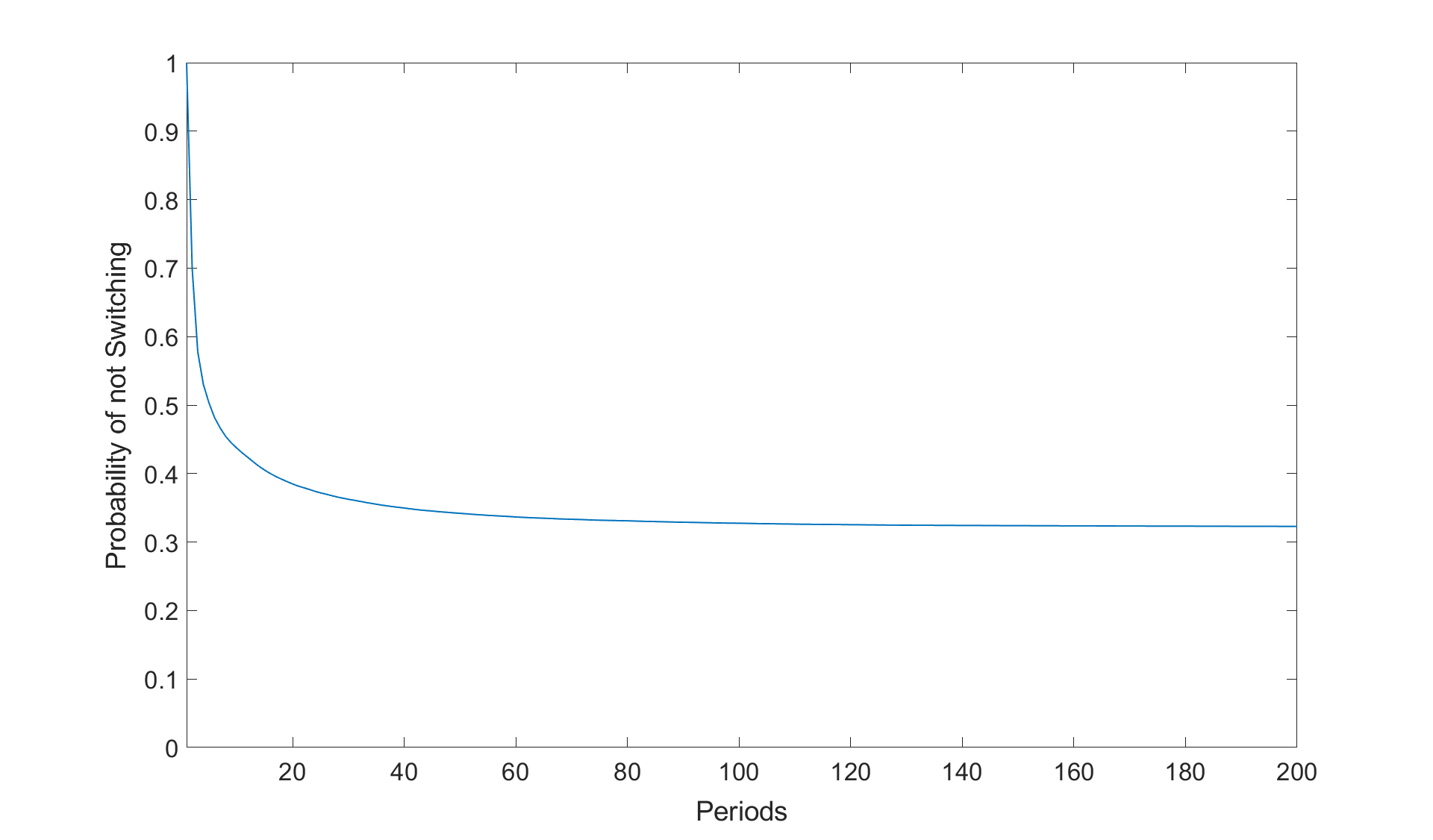}
    \caption{Simulation result on probability of the buyer using $P$ at each period ($l(A_t^{\alpha}(P)$) in the Seller-Buyer Example. The switching threshold $\alpha =1.39$, and the number of repetitions is 100,000.}
    \label{fig:simul1}
\end{figure}

\indent We ran simulations of 200 periods of history realizations under the true data-generating process $P$ with 100,000 repetitions, focusing on the changes of $l(A_t^{\alpha})(P)$ over periods. Figure~\ref{fig:simul1} shows how $l(A_t^{\alpha})(P)$ changes when the switching threshold $\alpha$ is 1.39. It initially decreases rapidly and converges to a value of around 0.32. Since it takes less than 0.5 after several periods, a patient seller would not use $P$ to persuade the buyer.

\begin{figure}[htb!]
    \centering
    \includegraphics[width=1\textwidth]{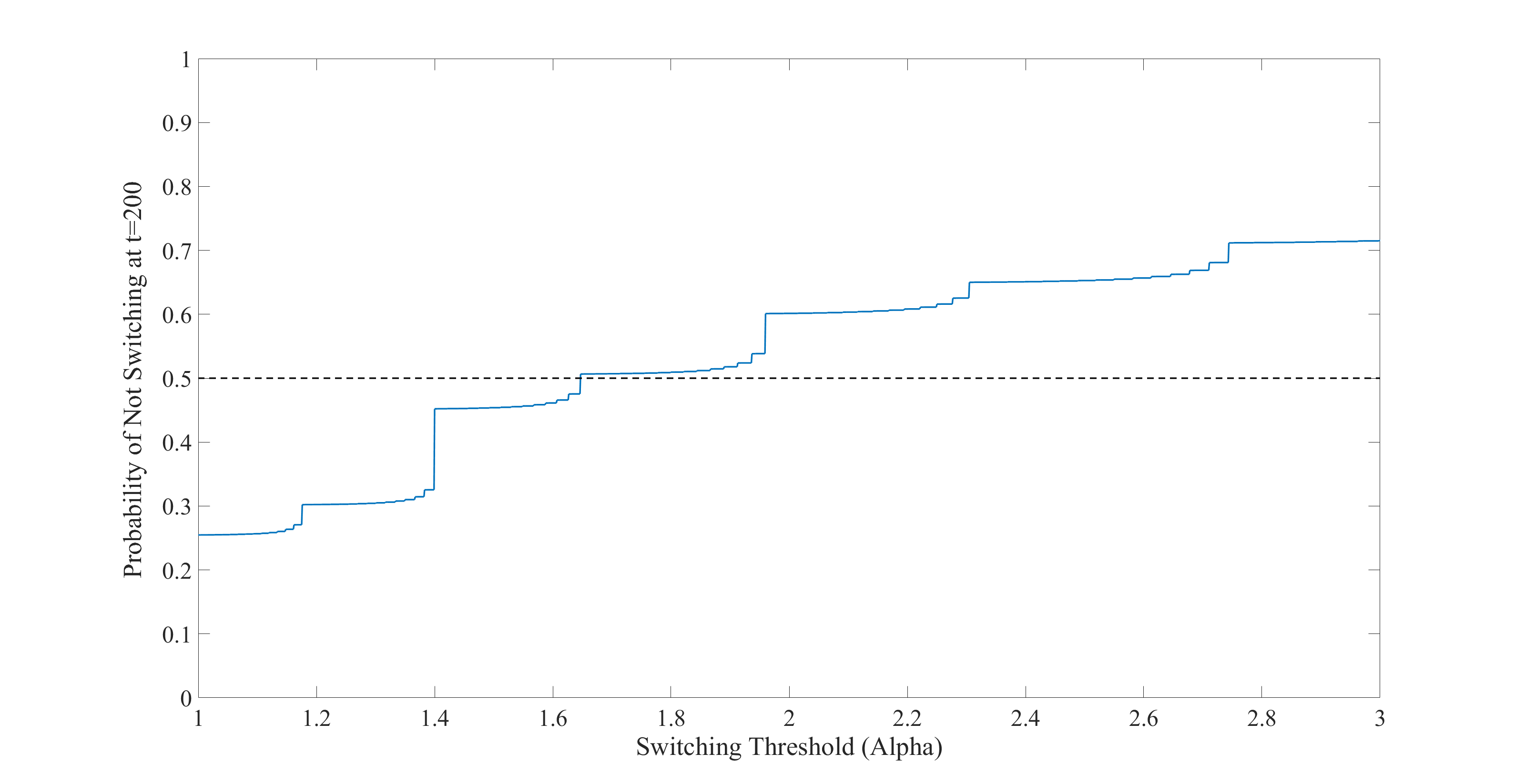}
    \caption{Simulation result on probability of the buyer using $P$ at period $t=200$ ($l(A_{200}(P)$) in the Seller-Buyer Example. The switching threshold $\alpha$ varies from 1 to 3, and the number of repetitions is 100,000.}
    \label{fig:simul2}
\end{figure}

\indent Then, we varied the value of $\alpha$ from 1 to 3 and derived $l(A_{200}(P))$ under each switching threshold. Figure~\ref{fig:simul2} presents the result. $l(A_{200}(P))$ increases with the threshold $\alpha$. Consistent with \textbf{Proposition~\ref{prop:seller}}, there exists $\hat{\alpha}$ between 1.6 and 1.7 such that $P$ cannot be optimal for the patient seller in the long run since it is beaten by full disclosure $P_f$. However, as $\alpha$ increases, the patient seller will turn toward $P$.

\indent Even if full disclosure beats the BP-optimal $P$, it does not assure that full disclosure $P_f$ maximizes the seller's lifetime expected utility. An information structure $P_{\epsilon}$ between $P$ and $P_f$ (more informative than $P$ but not fully informative) may yield the higher lifetime expected utility for the Sender. Consider the following information structure $P_{\epsilon}$:
\begin{displaymath}
    P_{\epsilon}= \ \bordermatrix{
    & hh & h     & l    \cr
    H  & \epsilon  & 1 - \epsilon    & 0     \cr
    L  & 0   & \frac{3}{7} (1 - \epsilon)    & \frac{4}{7} + \frac{3}{7}\epsilon    \cr
            }.
\end{displaymath}
The alternative would be
\begin{displaymath}
    \hat{P}_{\epsilon}= \ \bordermatrix{
    & hh & h     & l    \cr
    H  & 0.3\epsilon  & 0.6(1 - \epsilon)    & 0.4 + 0.3\epsilon     \cr
    L  & 0.3\epsilon   & 0.6(1 - \epsilon)    & 0.4 + 0.3\epsilon    \cr
            }.
\end{displaymath}
When the perceived information structure is $P_{\epsilon}$, the observations that the buyer anticipates are $(hh, H)$, $(hh, L)$, $(h,H)$, $(h,L)$, and $l$. The likelihoods of those observations under each information structure are presented in Table~\ref{tab:appen1}.
\begin{table}[!ht]
    \centering
    \begin{threeparttable}
    \begin{tabular}{  C{7cm} | C{2.5cm}  C{2.5cm}  } 
        \hline
        Observation & $P_{\epsilon}$ & $\hat{P}_{\epsilon}$ \\ 
        \hline
        $(hh,H)$ & $0.3\epsilon$ & $0.09\epsilon$ \\
        $(hh,L)$ & 0             & $0.21\epsilon$ \\
        $(h,H)$  & $0.3- 0.3\epsilon$ & $0.18 -0.18\epsilon$ \\
        $(h,L)$  & $0.3- 0.3\epsilon$ & $0.42- 0.42\epsilon$ \\
        $l$      & $0.4 +0.3\epsilon$ & $0.4 + 0.3\epsilon$ \\ 
        \hline
    \end{tabular}
    \begin{tablenotes}
            \item \footnotesize{\textit{Notes}: $P_{\epsilon}$ is a more informative information structure than the BP-optimal information structure $P$, and $\hat{P}$ is a no-information structure generating the same signal distribution as $P_{\epsilon}$. The observations are what the buyer anticipates when her perceived information structure is $P_{\epsilon}$.}
    \end{tablenotes}
    \end{threeparttable}
    \caption{Likelihoods of Observations under Announced and Alternative Information Structures in the Seller-Buyer Example when the seller provides more information to the buyer}
    \label{tab:appen1}
\end{table}

\indent There is a trade-off for providing more information to the Receiver. The cost of using $P_{\epsilon}$ instead of $P$ reduces the period expected utility, given that the buyer still adopts the seller's announcement. At the same time, however, the seller's lifetime expected utility can increase from a higher $l(A_t^{\alpha})$. Note that $(hh, L)$ never occurs since the true information structure is $P_{\epsilon}$, and $(h, L)$ is the only observation that induces the buyer's switch. Its likelihood decreases by $0.3\epsilon$. Moreover, the Bayes factor decreases significantly after the realization of $(hh, H)$, which has a positive effect on the seller's lifetime expected utility.

\begin{figure}[htb!]
    \centering
    \includegraphics[width=1\textwidth]{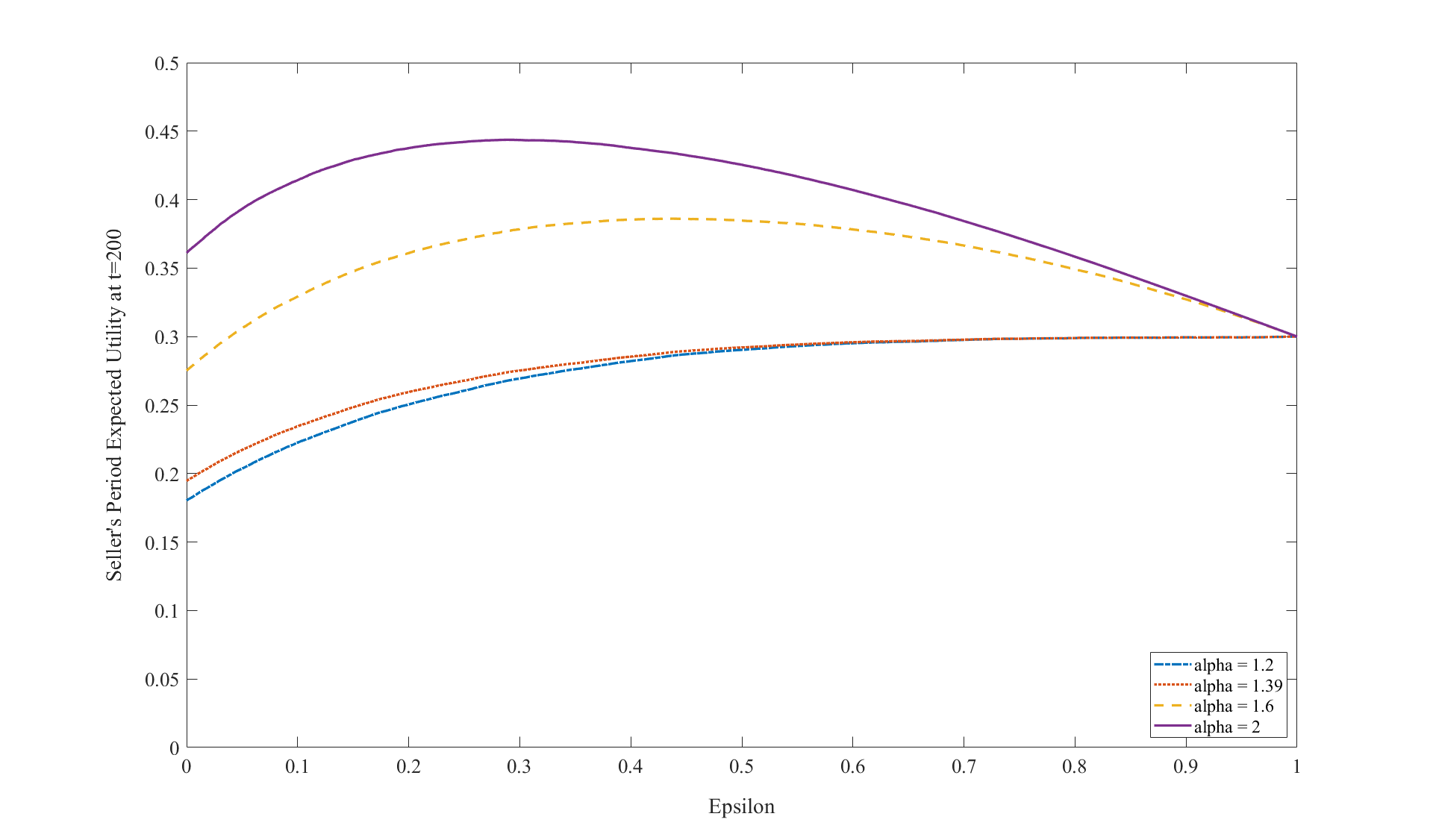}
    \caption{Simulation result on the seller's period expected utility at period $t=200$ ($l(A_{200}(P)$) using $P_{\epsilon}$ in the Seller-Buyer Example. $\epsilon$ varies from 0 to 1, and the number of repetitions is 100,000. $\alpha$ takes values of 1.2, 1.39, 1.6, and 2.}
    \label{fig:simul3}
\end{figure}

\indent The natural question arises: what would be the optimal level of $\epsilon$, i.e., what would be the optimal level of information that the seller should provide to the buyer? Our simulation results suggest a potential answer to this question when the seller is patient. Figure~\ref{fig:simul3} displays the simulation of the seller's period expected utility at $t=200$ varying $\epsilon$. It demonstrates that the patient seller can be better off by providing more information to the buyer as $\alpha$ decreases. When $\alpha$ goes down to 1, the optimal information structure moves to full disclosure, while the seller's optimal choice converges toward the BP-optimal one as $\alpha$ diverges. Also, it shows that his expected utility increases with higher $\alpha$ since he can entertain greater period expected utility with higher probability.

\end{document}